\documentclass[reqno,12pt]{amsart} %{article}
\usepackage{epsf}\usepackage{amssymb} \usepackage{cite}
\usepackage{amsthm}\usepackage{amscd}\usepackage{amsfonts}\usepackage{amsmath}

\newcommand{\be}{\begin{equation}}
\newcommand{\ee}{\end{equation}}
\newcommand{\bea}{\begin{eqnarray}}
\newcommand{\eea}{\end{eqnarray}}

\usepackage{hyperref}
\def\stackreb#1#2{\ \mathrel{\mathop{#1}\limits_{#2}}}

\newcommand{\CC}{\mathbb C}

\newcommand{\R}{\mathbb R}
\newcommand{\Z}{\mathbb Z}

\newcommand{\T}{\mathbb T}

\newcommand{\ve}{\varepsilon}

\begin{document}

\title[Complex hypergeometric functions and integrable many-body problems]
{Complex hypergeometric functions and   \\ integrable many-body problems}

\author{G. A. Sarkissian  and V. P. Spiridonov }

\address{
Sarkissian:  Laboratory of Theoretical Physics, JINR, Dubna,   Russia
and  Yerevan Physics Institute, Alikhanian Brothers 2, 0036 Yerevan, Armenia
\hfill \break \indent
Spiridonov: Laboratory of Theoretical Physics, JINR, Dubna,    Russia  and  National Research University Higher School of Economics, Moscow, Russia}

\begin{abstract}
General reduction of the elliptic hypergeometric equation to the level of complex hypergeometric functions
is described. The derived equation is generalized to the Hamiltonian eigenvalue problem for new rational integrable
$N$-body systems emerging from particular degenerations of the elliptic Ruijsenaars and van Diejen  models.
\end{abstract}

\maketitle

\tableofcontents

\section{Introduction}

A classical integrable $N$-body system representing a relativistic generalization of the Calogero
and Sutherland models
has been suggested by Ruijsenaars and Schneider \cite{rel}. Its quantization associated with the  finite-difference
operators and a further extension to the elliptic interaction potential were considered by Ruijsenaars \cite{R87}.
The top model of such type was constructed by van Diejen \cite{vD}. The complete integrability of this model
was established by Komori and Hikami \cite{KH}. The models of \cite{rel,R87} are associated with the root system $A_N$, whereas the Hamiltonian of  \cite{vD} is invariant under the action of Weyl group for the root system $C_N$.

Elliptic functions can be sequentially degenerated to the trigonometric or hyperbolic functions and further
down to the rational functions. This hierarchy is present in the integrable models as well. From the special functions
point of view solutions of differential and, more generally, finite-difference equations with
the coefficients described by rational, trigonometric/hyperbolic or elliptic functions are related to the
plain hypergeometric functions---the Euler-Gauss $_2F_1$-function and its higher order and multivariable
generalizations, the $q$-hypergeometric series and integrals together with their hyperbolic analogues,
and, finally, the elliptic hypergeometric functions \cite{spi:thesis}, respectively. The most important
application of the elliptic hypergeometric integrals has been found in the quantum field theory --- they
describe superconformal indices of four-dimensional supersymmetric gauge field theories \cite{DO,SV}.
Therefore any new result on these functions is important for the quantum field theory.

Classical and quantum integrable many-body problems have found many applications in physics.
In the simplest setting they describe $N$ non-relativistic particles with the specific
pairwise interaction lying on a line or circle with external potentials of particular types.
These systems admit exact description of their dynamics, in particular, eigenfunctions of
the corresponding Hamiltonians are described
by some well understood special functions. Substantially less is known for the elliptic potential interaction
cases and the relativistic models, despite of the presence of a complete number of integrals of motion.
Their further investigation is an interesting and active field of research due to the applications
in quantum field theory (in particular, to superconformal theories), solid state physics, physics
of solitons, etc.

The described hierarchy of interaction potentials from elliptic to the rational ones is not fully explored yet.
Already long ago it was recognized that there are complex extensions
of the plain hypergeometric functions associated with the representation theory of $SL(2,\CC)$ group
\cite{GGV,Naimark}. Such functions are defined either as (multiple) integrals over the complex plane
or by infinite bilateral sums of the Mellin-Barnes type integrals. By now they have found
various applications in different fields of theoretical and mathematical physics, see a list
of relevant references in \cite{DM2019,MN,Sarkissian:2020ipg}. In particular, they play an
important role in the two-dimensional conformal field theory and non-compact Heisenberg spin chains
with applications to the quantum chromodynamics \cite{DKM}. However, until recently the relation of
complex hypergeometric functions to elliptic hypergeometric functions was not known and,
moreover, they were not considered in the context of completely integrable quantum many-body systems,
which is the main purpose of the present work. Namely, we show that there exists a new rational
degeneration of the Ruijsenaars and van Diejen elliptic integrable systems appearing from a particular
degeneration of the Faddeev modular dilogarithm \cite{Fad94}
to the ``complex'' gamma function rigorously established
in \cite{Sarkissian:2020ipg}. An interesting fact is that the orthogonality measure for eigenfunctions
of the corresponding rational limit of a simplified van Diejen model is described by a complex hypergeometric integral
in the Mellin-Barnes representation, which contains as a special subcase a complex extension of
the celebrated Selberg integral \cite{FW}.

The key elliptic Ruijsenaars model Hamiltonian has the form  \cite{R87}
\begin{equation}
\mathcal{H}_R=\sum_{j=1}^N\prod_{k=1,\neq j}^N \frac{\theta(tz_jz_k^{-1};p)}{\theta(z_jz_k^{-1};p)}\,T_j,
\label{HR}\end{equation}
where  $z_j, t, q \in \CC^\times, p \in\CC, |p|<1$, and $T_j$ are the $q$-shift operators
$$
T_j\psi(z_1,\ldots,z_j,\ldots,z_N)=\psi(z_1,\ldots,qz_j,\ldots,z_N).
$$
We denoted as $\theta(z;p)$ the $p$-theta function
$$
\theta(z;p)=(z;p)_\infty(pz^{-1};p)_\infty, \qquad (z;p)_\infty=\prod_{k=0}^\infty (1-zp^k).
$$
The standard Jacobi theta functions with characteristics have the form
$$
\theta_{ab}(u)=\sum_{\ell\in \mathbb Z}e^{\pi \textup{i} \tau (\ell+a/2)^2}
e^{2\pi \textup{i} \ell (u+b/2)}, \quad a, b=0, 1.
$$
Another popular notation for these functions is
$$
\theta_1(u)=-\theta_{11}(u),\quad \theta_2(u)=\theta_{10}(u), \quad
\theta_3(u)=\theta_{00}(u), \quad \theta_4(u)=\theta_{01}(u).
$$
Using the Jacobi triple product identity, one finds the relation of $\theta(z;p)$ function described above
to $\theta_1(u)$:
\begin{equation}
\theta_1(u)=\textup{i}p^{1/8}e^{-\pi \textup{i} u}(p;p)_\infty\theta(e^{2\pi \textup{i}u};p),
\qquad p=e^{2\pi \textup{i}\tau}.
\label{theta1}\end{equation}

The original elliptic van Diejen model was formulated with the help of the Weierstrass sigma function \cite{vD}.
We describe its Hamiltonian in terms of the Jacobi theta functions (as in \cite{KH}):
$$
\mathcal{H}_{vD}=\sum_{j=1}^N\Big(b_j(\underline{u})e^{2\gamma\partial_{u_j}}
+b_j(-\underline{u})e^{-2\gamma\partial_{u_j}}\Big)+ v(\underline{u}),
$$
where shifting operators act on functions of $u_j$-variables as
$e^{2\gamma\partial_{u_j}}f(\ldots, u_j, \ldots)=f(\ldots, u_j+2\gamma,\ldots)$,
the coefficients $b_j(\underline{u})$ have the form
$$
b_j(\underline{u})=\prod_{k=1, \neq j}^N\frac{\theta_1(u_j\pm u_k+\mu)}{\theta_1(u_j\pm u_k)}
\prod_{r=0}^3\frac{\theta_{r+1}(u_j+\mu_r)\theta_{r+1}(u_j+\gamma+\mu_r')}
{\theta_{r+1}(u_j)\theta_{r+1}(u_j+\gamma)},
$$
and the potential $v(\underline{u})$  is an elliptic function of $u_j$ having a rather lengthy representation
\begin{eqnarray*} &&
v(\underline{u})=\left(\frac{\pi}{\theta'_1(0)}\right)^2\frac{2}{\theta_1(\mu)\theta_1(\mu-2\gamma)}
\Big(\prod_{r=0}^3\theta_{r+1}(\mu_r-\gamma)\theta_{r+1}(\mu_r')
\prod_{j=1}^N\frac{\theta_1(\pm u_j+\mu-\gamma)}{\theta_1(\pm u_j-\gamma)}
\\ && \makebox[-1em]{}
+\theta_1(\mu_1-\gamma)\theta_1(\mu_1')\theta_2(\mu_0-\gamma)\theta_2(\mu_0')
\theta_3(\mu_3-\gamma)\theta_3(\mu_3')\theta_4(\mu_2-\gamma)\theta_4(\mu_2')
\prod_{j=1}^N\frac{\theta_2(\pm u_j+\mu-\gamma)}{\theta_2(\pm u_j-\gamma)}
\\ && \makebox[-1em]{}
+\theta_1(\mu_2-\gamma)\theta_1(\mu_2')\theta_2(\mu_3-\gamma)\theta_2(\mu_3')
\theta_3(\mu_0-\gamma)\theta_3(\mu_0')\theta_4(\mu_1-\gamma)\theta_4(\mu_1')
\prod_{j=1}^N\frac{\theta_3(\pm u_j+\mu-\gamma)}{\theta_3(\pm u_j-\gamma)}
\\ && \makebox[-1em]{}
+\theta_1(\mu_3-\gamma)\theta_1(\mu_3')\theta_2(\mu_2-\gamma)\theta_2(\mu_2')
\theta_3(\mu_1-\gamma)\theta_3(\mu_1')\theta_4(\mu_0-\gamma)\theta_4(\mu_0')
\prod_{j=1}^N\frac{\theta_4(\pm u_j+\mu-\gamma)}{\theta_4(\pm u_j-\gamma)}\Big),
\end{eqnarray*}
where we denoted $\theta_r(\mu\pm u):= \theta_r(\mu+ u)\theta_r(\mu- u)$.

Substituting relation \eqref{theta1} and similar representations for $\theta_{2,3,4}(u)$-functions
into the coefficients $b_j$ we obtain
$$
b_j(\underline{u})=e^{-2\pi\textup{i}\mu(N-1)-\pi\textup{i}(\mu_0+\mu_0'+\mu_1+\mu_1')} A_j(\underline{z}),
$$
where
\be
 A_j(\underline{z})=\frac{\prod_{m=1}^8\theta(t_mz_j;p)}{\theta(z_j^2,qz_j^2;p)}
\prod_{k=1\atop k \ne j}^N
\frac{\theta(tz_jz_k,tz_jz_k^{-1};p)}{\theta(z_jz_k,z_jz_k^{-1};p)}
\label{Aj}\ee
with the notation
$$
z_j=e^{2\pi\textup{i}u_j},\; t=e^{2\pi\textup{i}\mu},\; t_1=-e^{2\pi\textup{i}\mu_1},\;
t_2=-p^{1/2}e^{2\pi\textup{i}\mu_2},\; t_3=p^{1/2}e^{2\pi\textup{i}\mu_3},\;
t_4=e^{2\pi\textup{i}\mu_0},
$$
$$
t_5=q^{1/2}e^{2\pi\textup{i}\mu_0'}, \;
t_6=-q^{1/2}e^{2\pi\textup{i}\mu_1'}, \;
t_7=-q^{1/2}p^{1/2}e^{2\pi\textup{i}\mu_2'}, \;
t_8=q^{1/2}p^{1/2}e^{2\pi\textup{i}\mu_3'}, \;
q=e^{4\pi\textup{i}\gamma}.
$$
We use also the convention $\theta(t_1, \ldots, t_k;p):=\theta(t_1;p)\ldots \theta(t_k;p)$.
In the transition from $b_j$ to $A_j$-coefficients we employed the formula of
doubling of the argument for $p$-theta function, $\theta(z,-z,p^{1/2}z,-p^{1/2}z;p)=\theta(z^2;p)$.

Under the balancing condition
\begin{equation}
2(N-1)\mu+\sum_{r=0}^3(\mu_r+\mu_r')=0, \quad \textup{or} \quad
t^{2N-2}\prod_{m=1}^8t_m=p^2q^2,
\label{balance1}\end{equation}
the coefficients $A_j(\underline{z})$ become $p$-elliptic functions of all variables $z_k$,
$$
A_j(\ldots,p z_k,\ldots)=A_j(\underline{z}),
$$
which follows from basic properties of the $\theta$-function
$$
\theta(pz;p)=\theta(z^{-1};p)=-z^{-1}\theta(z;p).
$$
Moreover, as shown in \cite{vD,R3}, under the balancing condition the combination of functions
$$
\sum_{j=1}^N\big(b_j(\underline{u})+b_j(-\underline{u})\big)+v(\underline{u})
$$
does not depend on $u_j$, i.e. the potential $v(\underline{u})$ factorizes to
a very simple combination of the $b_j$-functions up to some additive constant.
Using this fact, we can take in this case
as the Hamiltonian of the van Diejen model a much simpler factorized operator
\be
\mathcal{H}_{vD}=\sum_{j=1}^N\Big(A_j(\underline{z})(T_j-1)
+A_j(\underline{z}^{-1})(T_j^{-1}-1)\Big),
\label{H}\ee
where we replaced the shifting operators $e^{2\gamma\partial_{u_j}}$ by $q$-shift operators $T_j$ defined in \eqref{HR}
(note that in the operator $\mathcal{H}_{R}$ \eqref{HR} one can replace $\theta(z;p)$ functions by $\theta_1(u)$ ones up
to some simple multiplicative constant).

The basic problem for stationary quantum mechanical systems
consists in solving the Schr\"odinger equation, or the eigenvalue problem
for the Hamiltonian
$$
\mathcal{ H}_{vD}\psi(\underline{z})=\lambda \psi(\underline{z}).
$$
Explicitly, the one-particle model, $N=1$, yields the following second-order analytic $q$-difference equation
\begin{equation}
\frac{\prod_{j=1}^8\theta(t_jz;p)}{\theta(z^2,qz^2;p)} (\psi(qz)-\psi(z))
+ \frac{\prod_{j=1}^8\theta(t_jz^{-1};p)}{\theta(z^{-2},qz^{-2};p)}
(\psi(q^{-1}z)-\psi(z))=\lambda \psi(z).
\label{n1}\end{equation}

\section{Elliptic hypergeometric equation}

Elliptic hypergeometric integrals discovered in \cite{spi:umn} form the most complicated class of transcendental special
functions of hypergeometric type. In \cite{spi:theta} the second author has considered
the following function
\be
V(t_1,\dots,t_8;p,q)=\frac{(p;p)_\infty (q;q)_\infty}{4\pi\textup{i}}\int_\T\frac{\prod_{j=1}^8\Gamma(t_jx^{\pm 1};p,q)}
{\Gamma(x^{\pm2};p,q)}\frac{dx}{x},
\label{V}\ee
where variables $t_j\in\CC^\times$  are subject to the constraints $|t_j|<1$ and the balancing condition
\be
\prod_{j=1}^8t_j=p^2q^2
\label{balance}\ee
(these $t_j$ should not be confused with the parameters entering \eqref{Aj}).
In \eqref{V} $\T$ is the unit circle of positive orientation and
$\Gamma(z;p,q)$ is the standard elliptic gamma function
$$
\Gamma(z;p,q)=\prod_{j,k=0}^\infty\frac{1-z^{-1}p^{j+1}q^{k+1}}{1-zp^jq^k}, \quad |p|, |q|<1,\; z\in \CC^\times,
$$
with the convention
$$
\Gamma(tz^{\pm 1};p,q)=\Gamma(tz;p,q)\Gamma(tz^{-1};p,q),\quad
\Gamma(t_1,\ldots,t_m;p,q)=\Gamma(t_1;p,q)\cdots \Gamma(t_m;p,q).
$$
The constraints $|t_j|<1$ guarantee that the integration contour $\T$ in \eqref{V} separates geometric
progression sequences of poles of
the integrand function converging to zero from those diverging to infinity. For other
domains of values of $t_j$ the $V$-function is defined by analytical continuation
(e.g., by deforming the integration contour).

The function \eqref{V} emerged from the analysis of symmetry transformations for multidimensional
elliptic hypergeometric integrals. In particular, from the exact elliptic beta integral evaluation
formula \cite{spi:umn}
\begin{equation}\makebox[-2em]{}
\frac{(p;p)_\infty (q;q)_\infty}{4\pi\textup{i}}\int_\T\frac{\prod_{j=1}^6\Gamma(t_jx^{\pm 1};p,q)}
{\Gamma(x^{\pm2};p,q)}\frac{dx}{x}=\prod_{1\leq j<k \leq 6}\Gamma(t_jt_k;p,q),
\label{ellbeta}\end{equation}
where $\prod_{j=1}^6t_j=pq, \; |t_j|<1$,
a symmetry transformation of the $V$-function was established in \cite{spi:theta},
which extended the obvious $S_8$ symmetry group of permutations of parameters to $W(E_7)$---the Weyl group
for the exceptional root system $E_7$.  The identity \eqref{ellbeta} was proved with the help of the contiguous
relations following from the addition formula for theta functions and a novel analytical continuation argument.
As shown in  \cite{spi:thesis,spi:tmf}, a combination of the latter contiguous relations and $V$-function
symmetries yields the following elliptic hypergeometric equation
\be\label{elldif}
B(\underline{t};q;p)(U(qt_6,q^{-1}t_7)-U(\underline{t}))+(t_6\leftrightarrow t_7)+U(\underline{t})=0,
\ee
where
\be
U(\underline{t})=\frac{V(t_1,\ldots,t_8;p,q)}{\Gamma(t_6 t_8^{\pm 1}; p,q)\Gamma(t_7 t_8^{\pm 1}; p,q)}
\label{f2}\ee
and
\bea\label{ba}
B(\underline{t};q;p)=\frac{\theta\left(\frac{t_6}{qt_8};p\right)\theta\left(t_6 t_8; p\right)
\theta\left(\frac{t_8}{t_6};p\right)}{
\theta\left(\frac{t_6}{t_7}; p\right)\theta\left(\frac{t_7}{qt_6};p\right)
\theta\left(\frac{t_7t_6}{q};p\right)}
\prod_{k=1}^5\frac{\theta\left(\frac{t_7t_k}{q};p\right)}{\theta(t_8t_k;p)}.
\eea

In \eqref{elldif} $U(qt_j,q^{-1}t_k)$ represents the function $U(\underline{t})$ in which parameters $t_j$ and $t_k$
were replaced by $qt_j$ and $q^{-1}t_k$, respectively. Also, the notation $(t_6\leftrightarrow t_7)$ means
that there stands the preceding expression with the permuted parameters $t_6$ and $t_7$.

For a special choice $t_jt_k=q^{-n}p^{-m},\,  n,m\in\mathbb{Z}_{\geq 0}$ for a pair of different parameters,
the $V$-function reduces to a product of two terminating $_{12}V_{11}$ elliptic hypergeometric
series \cite{spi:theta}. Each of these series define an infinite-dimensional set of rational functions biorthogonal
with respect to the measure defined by the elliptic beta integral \cite{spi:umn}. A finite-dimensional
reduction, reached by a discretization of one more parameter, leads to the biorthogonal rational
functions discovered in \cite{SZ2001,SZ2000}. In this case the elliptic hypergeometric equation \eqref{elldif}
is reduced to the three-term recurrence relation for the corresponding discrete biorthogonal rational
functions which was also derived in \cite{SZ2001,SZ2000}.

After denoting $t_6=cz,\, t_7=cz^{-1}$, equation \eqref{elldif} takes the form of a $q$-difference
equation of the second order. Introducing new variables
\be
\ve_k=\frac{q}{ct_k},\; k=1,\ldots, 5,\quad \ve_6=p^4ct_8,
\quad \ve_7=\frac{c}{qt_8},   \quad \ve_8=\frac{c}{t_8}, \quad \prod_{k=1}^8 \ve_k=p^2q^2,
\label{change}\ee
where $\ve_8=q\ve_7$ and
$$
c^2=t_6t_7=\frac{\ve_6\ve_8}{p^4},
$$
equation \eqref{elldif} can be rewritten in the following symmetric form
\begin{eqnarray}\nonumber &&
\frac{\prod_{j=1}^8\theta(\ve_jz;p)}{\theta(z^2,qz^2;p)} (\psi(qz)-\psi(z))
+ \frac{\prod_{j=1}^8\theta(\ve_jz^{-1};p)}{\theta(z^{-2},qz^{-2};p)}(\psi(q^{-1}z)-\psi(z))
\\ && \makebox[6em]{}
+\prod_{k=1}^6\theta\left(\frac{\ve_k \ve_8}{q};p\right)\psi(z)=0.
\label{n1'}\end{eqnarray}

Now it is clearly seen \cite{spi:thesis,spi:tmf} that the elliptic
hypergeometric equation \eqref{elldif}  coincides
with the eigenvalue problem \eqref{n1} under the following constraints
(analogous observation was done by Komori).
First, one has to redenote parameters $t_j$ in \eqref{n1}
as $\ve_j$, then demand that $\ve_7=\ve_8/q$ and choose a special eigenvalue
\begin{equation}
\lambda= -\prod_{k=1}^6\theta\left(\frac{\ve_k\ve_8}{q};p\right).
\label{eigenvalue}\end{equation}
As a result, an explicit solution of the reduced equation \eqref{n1} is given by the function
\be
\psi(z)=\frac{V(q/c\ve_1,\ldots,q/c\ve_5,cz,c/z,c/\ve_8;p,q)}
{\Gamma(\ve_8z^{\pm1},c^2 z^{\pm1}/\ve_8;p,q)},
\label{f1}\ee
which is the very first explicit solution of eigenvalue problems for the relativistic
elliptic Ruijsenaars type models. After the reduction of the $V$-function to biorthogonal
rational functions of \cite{SZ2001,SZ2000} and \cite{spi:theta}  these discrete sets of functions
also represent particular eigenfunctions of the van Diejen Hamiltonian \eqref{H}.

Of course, solution \eqref{f1} is defined up to the multiplication by an arbitrary
elliptic function $\varphi(qz)=\varphi(z)$ (a quasiconstant). The second independent solution can be obtained
in several different ways \cite{spi:thesis,spi:tmf} by applying to \eqref{f1} symmetries of the elliptic hypergeometric
equation \eqref{n1}, which are not symmetries of the function \eqref{f1}.
For instance, the potential $B(\underline{t};q;p)$ in equation \eqref{elldif} is an elliptic function of all its variables
$$
B(\ldots, p^{n_j}t_j,\ldots ;p^mq;p)=B(\underline{t};q;p), \quad \sum_{j=1}^8n_j=2m,\quad n_j, m\in\Z,
$$
which means that equation \eqref{n1'} does not change its form if we scale $\ve_j\to p^{n_j}\ve_j,\, n_7=n_8$
and $z\to p^{n_0}z$ with arbitrary integers $n_j$, $\sum_{j=1}^8n_j=0$
(we cannot scale $q$ since it changes the form of the equation).
Therefore other solutions of equation \eqref{n1'} are obtained by multiplying $\ve_j$ and $z$ by
the indicated powers of $p$.

Note that equation \eqref{n1'} is explicitly symmetric in parameters $\ve_1,\ldots, \ve_6$ whereas
its solution \eqref{f1} is symmetric only in $\ve_1,\ldots, \ve_5$. Therefore, simple permutations
of $\ve_6$ with $\ve_1,\ldots,\ve_5$ in \eqref{f1} yield new solutions.
Another important option consists in employing the modified elliptic gamma
function \cite{spi:theta} or, alternatively, in the use of the modular
transformation properties of theta functions. Namely, $\theta(z;p)$-function obeys the following
$SL(2,\Z)$-transformation rule
\begin{equation}
\theta\left(e^{-2\pi \textup{i}\frac{u}{\omega_1}};
e^{-2\pi \textup{i}\frac{\omega_2}{\omega_1}}\right)
=e^{\pi \textup{i}B_{2,2}(u|\omega_1,\omega_2)} \theta\left(e^{2\pi \textup{i}\frac{u}{\omega_2}};
e^{2\pi \textup{i}\frac{\omega_1}{\omega_2}}\right),
\label{mod-theta}\end{equation}
where $B_{2,2}(u|\omega_1,\omega_2)$ is the second order Bernoulli polynomial
$$
B_{2,2}(u|\omega_1,\omega_2)=\frac{1}{\omega_1\omega_2}\left(u^2-(\omega_1+\omega_2)u+\frac{\omega_1^2
+\omega_2^2}{6}+ \frac{\omega_1\omega_2}{2}\right).
$$
 After parametrization
\begin{equation}
q=e^{2\pi \textup{i} \frac{\omega_1}{\omega_2}}, \quad
p=e^{2\pi \textup{i} \frac{\omega_3}{\omega_2}}, \quad
\ve_j=e^{2\pi \textup{i} \frac{g_j}{\omega_2}}, \quad
z=e^{2\pi \textup{i} \frac{u}{\omega_2}},
\label{sl3z}\end{equation}
equation \eqref{n1'} becomes invariant with respect to the modular transformation
$(\omega_2,\omega_3)\to (-\omega_3,\omega_2)$ for fixed parameters $g_j$.
More precisely, after application of the transformation \eqref{mod-theta} in appropriate
parametrization all three parts of equation \eqref{n1'} acquire
equal multiplicative factors (in terms of the equation \eqref{elldif} the potential $B$ is
simply modular invariant). Note that the $z\to qz$ transformation corresponds to the shift
$u\to u +\omega_1$, which is independent of $\omega_2$ and $\omega_3$.
 Therefore, if one applies the indicated parametrization
and modular transformation to the function $U(\underline{t})$, this also yields
a new independent solution of the elliptic hypergeometric function, which appears
to be well defined even in the case $|q|=1$ due to the properties of such a
parametrization \cite{spi:theta}.

Evidently, the indicated solutions of  the elliptic hypergeometric equation cannot all be independent.
An interesting open problem is to build an elliptic analogue of Kummer's 24 solutions to ordinary
hypergeometric equation and to construct elliptic analogues of linear relations between these dependent solutions.
The $V$-function symmetry group is very large and its application to the elliptic hypergeometric
equation \eqref{elldif} leads to a large number of different equations. It is necessary to describe all
possible minimal systems of equations which would fix the $V$-function uniquely similar to the choice
suggested in \cite{RS}.

An important point is that the $U$-function in \eqref{elldif} is symmetric in $p$ and $q$.
Therefore one has an additional analytic difference equation for it also having the form
of the elliptic hypergeometric equation:
 \be\label{elldif2}
B(\underline{t};p;q)(U(pt_6,p^{-1}t_7)-U(\underline{t}))+(t_6\leftrightarrow t_7)+U(\underline{t})=0.
\ee
However, for fixed parameters $\ve_k$ the function \eqref{f1} is not invariant under the permutation of $p$ with $q$.
Therefore, rewriting of \eqref{elldif2} in terms of the parameters $\ve_k$ defined in
\eqref{change} leads to the symmetric elliptic hypergeometric equation with a different choice of parameters
\begin{eqnarray}\nonumber &&
\frac{\prod_{j=1}^8\theta(\tilde \ve_jz;q)}{\theta(z^2,pz^2;q)} (\psi(pz)-\psi(z))
+ \frac{\prod_{j=1}^8\theta(\tilde \ve_jz^{-1};q)}{\theta(z^{-2},pz^{-2};q)}(\psi(p^{-1}z)-\psi(z))
\\ && \makebox[6em]{}
+\prod_{k=1}^6\theta\left(\frac{\tilde \ve_k\tilde \ve_8}{p};q\right)\psi(z)=0,
\label{n2} \\ &&
\tilde \ve_k=\frac{p}{q}\ve_k, \; k=1,\ldots,5, \quad \tilde \ve_6=\frac{q^4}{p^4} \ve_6,
\quad \tilde \ve_7=\frac{q}{p}\ve_7, \quad \tilde \ve_8=\ve_8, \quad \prod_{k=1}^8\tilde \ve_k=p^2q^2,
\nonumber\end{eqnarray}
which obviously does not coincide with the $p\leftrightarrow q$ permuted form of  equation
\eqref{n1'}. To illustrate the variety of possibilities, let us make the following remark.
As mentioned above,  modular transformation $(\omega_2, \omega_3)\to (-\omega_3, \omega_2)$ of $U(\underline{t})$
yields an independent solution of equation \eqref{n1'}. However, the second difference equation
satisfied by this transformed function is quite different from \eqref{n2}, which we do not present for brevity.

\section{Degeneration to the plain and complex hypergeometric functions}

Now we turn to the plain and complex hypergeometric functions emerging in particular rational degeneration
limits for the elliptic hypergeometric function \eqref{V}. Let us take the following parametrization
\begin{equation}
t_a=e^{-2\pi v g_a}, \qquad x=e^{-2\pi v z},\qquad p=e^{-2\pi v\omega_1}, \qquad q=e^{-2\pi v\omega_2},
\label{newpar}\end{equation}
where $\omega_1$ and $\omega_2$ should not be confused with the quasiperiods indicated in \eqref{sl3z}.
The limit  $v\to 0^+$ was formally applied for the first time to the elliptic beta integral in \cite{stok}.
As shown in \cite{rai:limits}, such a limit is uniform and can be rigorously used for degenerating the
elliptic hypergeometric integrals. So, the $V$-function has the following asymptotic expression
in the  limit $v\to 0^+$,
$$
V(\underline{t};p,q)\stackreb{=}{v\to 0^+} e^{\frac{5\pi}{12v}\left(\frac{1}{\omega_1}+\frac{1}{\omega_2}\right)}
\, I_h(\underline{g};\omega_1,\omega_2),
$$
where $I_h$ is a particular hyperbolic hypergeometric integral
\be\label{bi}
I_h(\underline{g};\omega_1,\omega_2)=\int_{-\textup{i}\infty}^{\textup{i}\infty}
\frac{
\prod_{j=1}^8\gamma^{(2)}(g_j\pm z;\omega_1,\omega_2)}{\gamma^{(2)}(\pm 2z;\omega_1,\omega_2)}
\frac{dz}{2\textup{i}\sqrt{\omega_1\omega_2}}
\ee
with the parameters $g_j$ satisfying the conditions Re$(g_j)>0$ and
\be\label{mu8}
\sum_{j=1}^8 g_j=2(\omega_1+\omega_2).
\ee
The integrand of \eqref{bi} is composed out of the Faddeev modular quantum dilogarithm \cite{Fad94,Fad95}
known also as the hyperbolic gamma function \cite{rui:first,RuijAW}
\begin{eqnarray} \nonumber &&
\gamma^{(2)}(u;\omega_1,\omega_2):=e^{-\frac{\pi\textup{i}}{2}
B_{2,2}(u;\omega_1,\omega_2) }
\frac{(\tilde { q }\, e^{2\pi \textup{i} \frac{u}{\omega_1}};\tilde { q })_\infty}
{(e^{2\pi \textup{i} \frac{u}{\omega_2}};{ q })_\infty}
\\ && \makebox[4em]{}
=\exp\left(-\frac{\pi\textup{i}}{2}
B_{2,2}(u;\omega_1,\omega_2) -\int_{\R+\textup{i}0}\frac{e^{ux}}
{(1-e^{\omega_1 x})(1-e^{\omega_2 x})}\frac{dx}{x}\right).
\label{int_rep}\end{eqnarray}
In \eqref{bi} we use the compact notation $\gamma^{(2)}(g\pm u;\mathbf{\omega})
:=\gamma^{(2)}(g+u;\mathbf{\omega})\gamma^{(2)}(g-u;\mathbf{\omega})$.
Also we have denoted
$$
{ q}=e^{2\pi\textup{i}\frac{\omega_1}{\omega_2}}, \qquad \tilde { q}
=e^{-2\pi\textup{i}\frac{\omega_2}{\omega_1}},
$$
where now $\omega_{1, 2}$ are the quasiperiods introduced in \eqref{newpar}.

Using the modular transformation rule \eqref{mod-theta} it is easy to derive
asymptotic expression for the theta function in the $v\to 0^+$ limit
$$
\theta(e^{-2\pi v u};e^{-2\pi v \omega_1})\stackreb{=}{v\to 0^+}
e^{-\frac{\pi}{6\omega_1 v}}2\sin\frac{\pi u}{\omega_1}.
$$
Now one can apply the described limit to the elliptic hypergeometric equation \eqref{elldif}
and obtain the relation \cite{BRS}
\be\label{br}
{\mathcal A}(\underline{g},\omega_2;\omega_1)(Y(g_6+\omega_2,g_7-\omega_2)-Y(\underline{g}))
+(g_6\leftrightarrow g_7)+Y(\underline{g})=0,
\ee
where we have denoted the potential as
\bea \nonumber &&
{\mathcal A}(\underline{g},\omega_2;\omega_1):=\frac{\sin{\frac{\pi}{\omega_1}}(g_6-g_8-\omega_2)\sin{\frac{\pi}{ \omega_1}}(g_6+g_8)\sin{\frac{\pi}{ \omega_1}}(g_8-g_6)}
{\sin{\frac{\pi}{ \omega_1}}(g_6-g_7)\sin{\frac{\pi}{ \omega_1}}(g_7-g_6-\omega_2)\sin{\frac{\pi}{ \omega_1}}(g_7+g_6-\omega_2)}
\\ && \makebox[8em]{} \times
\prod_{k=1}^5\frac{\sin{\frac{\pi}{ \omega_1}}(g_7+g_k-\omega_2)}{\sin{\frac{\pi}{ \omega_1}}(g_8+g_k)}
\label{ba1}\eea
and the solution as
\be
Y(\underline{g};\omega_1,\omega_2):=\frac{I_h(\underline{g};\omega_1,\omega_2)
}{ \gamma^{(2)}(g_6\pm g_8, g_7\pm g_8; \omega_1,\omega_2)}.
\ee
Because the function $Y(\underline{g};\omega_1,\omega_2)$ is symmetric in $\omega_1$ and $\omega_2$,
we have the second difference equation obtained by permutation of these quasiperiods (it is obtained
from equation \eqref{elldif2} in the corresponding limit)
\be\label{br2}
{\mathcal A}(\underline{g},\omega_1;\omega_2)(Y(g_6+\omega_1,g_7-\omega_1)-Y(u))+(g_6\leftrightarrow g_7)+Y(u)=0.
\ee
This is similar to the situation considered in \cite{RuijAW} for a simpler hyperbolic integral.

Let us describe the standard rational degeneration of this equation using the limit $\omega_1\to 0$ for fixed $\omega_2$, when ${ q } \to 1$ and $\tilde { q } \to 0$.
One can rewrite the infinite product  in \eqref{int_rep} in the form
$$
\gamma^{(2)}(\omega_1 x;\omega_1,\omega_2)=\Gamma_{ q }(x)
\frac{e^{-\frac{\pi\textup{i}}{2}B_{2,2}(\omega_1x;\mathbf{\omega})}}
{({ q };{ q })_\infty (1-{ q })^{1-x}} (\tilde { q }\, e^{2\pi\textup{i} x};\tilde { q })_\infty,
$$
where
$$
\Gamma_{ q }(x)=\frac{({ q };{ q })_\infty}{({ q }^x;{ q })_\infty}
(1-{ q })^{1-x}\stackreb{=}{{ q }\to 1}\Gamma(x)
$$
is the Jackson $q$-gamma function. Applying modular transformation property
of the Dedekind $\eta$-function
$$\eta(-1/\tau)=\sqrt{-\textup i\tau} \eta(\tau), \qquad
\eta(\tau)=e^{\frac{\pi \textup{i}\tau}{12}}(e^{2\pi \textup{i}\tau};e^{2\pi \textup{i}\tau})_\infty
$$
one can find the asymptotics
\be
\gamma^{(2)}(\omega_1 x;\omega)
\stackreb{=}{\omega_1\to 0} \frac{\Gamma(x)}{\sqrt{2\pi}}
\left(\frac{\omega_2}{2\pi\omega_1}\right)^{\frac{1}{2}-x}.
\label{as}\ee
This asymptotic relation was rigorously obtained for the first time in \cite{rui:first}
in a different way, where it was also shown that it is uniform on the compacta.

In order to clean the balancing condition in $I_h$ we shift $g_{a,b}\to g_{a,b}+\omega_2$
for some indices $a\neq b$, so that $\sum_{j=1}^8g_j=2\omega_1$. Now we apply the inversion formula
$$
\gamma^{(2)}(x;\omega)\gamma^{(2)}(\omega_1+\omega_2-x;\omega)=1
$$
in order to flip $\gamma^{(2)}(g_{a,b}+\omega_2 \pm z)$-functions from the integrand numerator to
the denominator. After that we set $g_k:=\omega_1 \alpha_k$, scale the integration variable
$z=\omega_1 u$ in \eqref{bi}, and come to the asymptotic relation
$$
I_h(\underline{g};\omega_1,\omega_2)\stackreb{=}{\omega_1\to 0}
\left(\frac{1}{2\pi}\sqrt{\frac{\omega_2}{\omega_1}}\right)^5
I_{r}(\underline{\alpha}),
$$
where $I_{r}$ is the following plain hypergeometric integral
\be
I_{r}(\underline{\alpha})=\frac{1}{4\pi\textup{i}}
\int_{-\textup{i}\infty}^{\textup{i}\infty}\frac{\prod_{j=1, \neq a, b}^8\Gamma(\alpha_j\pm u)}
{\Gamma(\pm 2u)\prod_{j=a,b}\Gamma(1-\alpha_j\pm u)}du,\quad \sum_{k=1}^8\alpha_k=2.
\label{Irat}\ee

The form of the difference equations for this function depends on the choice of $a$ and $b$.
Let us take $a=4,\, b=5$; then from \eqref{br2} we easily obtain the following
analytic difference equation
\be
{\mathcal B}(\underline{\alpha})({\mathcal I}_{r}(\alpha_6+1,\alpha_7-1)-{\mathcal I}_{r}(\underline{\alpha}))
+(\alpha_6 \leftrightarrow \alpha_7)+{\mathcal I}_{r}(\underline{\alpha}) =0,
\label{rateq1}\ee
where
\be
{\mathcal B}(\underline{\alpha})=\frac{(\alpha_6-\alpha_8-1)(\alpha_6+\alpha_8)(\alpha_8-\alpha_6)}
{(\alpha_6-\alpha_7)(\alpha_7-\alpha_6-1)(\alpha_7+\alpha_6-1)}
\prod_{k=1}^5\frac{\alpha_7+\alpha_k-1}{\alpha_8+\alpha_k}
\label{Bgen}\ee
and
\be
{\mathcal I}_{r}(\underline{\alpha})=\frac{I_{r}(\underline{\alpha})}
{\Gamma(\alpha_6\pm \alpha_8)\Gamma(\alpha_7\pm \alpha_8)}.
\label{solution1}\ee

As to equation \eqref{br}, in the limit $\omega_1\to 0$ it yields a curious relation between
two different integrals
\be
\frac{\sin\pi(\alpha_8\pm\alpha_6)\prod_{k=4}^5\sin\pi(\alpha_7+\alpha_k)}
{\sin\pi(\alpha_6-\alpha_7)\prod_{k=1}^3\sin\pi(\alpha_8+\alpha_k)}
(\tilde {\mathcal I}_{r}(\underline{\alpha})-{\mathcal I}_{r}(\underline{\alpha}))
+(\alpha_6\leftrightarrow \alpha_7)+{\mathcal I}_{r}(\underline{\alpha})=0,
\label{ratid1}\ee
where
\be
\tilde {\mathcal I}_{r}(\underline{\alpha})=\frac{1}{4\pi\textup{i}}
\frac{\Gamma(1-\alpha_6\pm\alpha_8)}{\Gamma(\alpha_7\pm\alpha_8)}
\int_{-\textup{i}\infty}^{\textup{i}\infty}\frac{\prod_{k=1}^3\Gamma(\alpha_k\pm u)
\prod_{k=7}^8\Gamma(\alpha_k\pm u)}{\Gamma(\pm 2u)\prod_{k=4}^6\Gamma(1-\alpha_k\pm u)}\, du.
\label{tildeI}\ee
Since the functions ${\mathcal I}_{r}(\underline{\alpha})$ and $\tilde {\mathcal I}_{r}(\underline{\alpha})$
are related to non-terminating very-well-poised 2-balanced $_9F_8$-series, it would be
interesting to rewrite identity \eqref{ratid1} in terms of hypergeometric series. We believe
that it yields a new plain hypergeometric series identity which was not considered before. Taking into account
the role played by such $_9F_8$-series in the arithmetics of zeta values\cite{MZ}, this
relation may be of a particular interest. We stress that this and other similar identities originate from three
fundamental symmetries present for the elliptic hypergeometric function \eqref{V}: the contiguous relations
stemming from the addition formula for the Jacobi theta functions, the Weyl group symmetry $W(E_7)$
and the $S_2$-permutation symmetry in basic parameters $p$ and $q$ \cite{spi:theta,spi:thesis}.

Next, we apply to the derived hyperbolic integral relations a particular rational degeneration limit
that leads to complex hypergeometric functions. Namely, applying the logic used in the described
derivation of the estimate \eqref{as}, one can rigorously establish \cite{Sarkissian:2020ipg}
in the limit $\omega_1/\omega_2\to -1$ the following asymptotic relation
for particular values of the argument of $\gamma^{(2)}$-function holding uniformly on the compacta
with exponentially small corrections
\be\label{gam2lim2}
\gamma^{(2)}(\textup{i}\sqrt{\omega_1\omega_2}(n+x\delta);\omega_1,\omega_2)
\stackreb{=}{\delta\to 0^+} e^{\frac{\pi \textup{i}}{2}n^2} (4\pi\delta)^{\textup{i}x-1}{\bf \Gamma}(x,n),
\quad \sqrt{\frac{\omega_1}{\omega_2}}=\textup{i}+\delta,
\ee
where $n\in \Z, \, x\in\CC$. In \eqref{gam2lim2} ${\bf \Gamma}(x,n)$ is the gamma function
over the field of complex numbers
\cite{GGV}
\begin{equation}
{\bf \Gamma}(x,n)=\frac{\Gamma(\frac{n+\textup{i}x}{2})}{\Gamma(1+\frac{n-\textup{i}x}{2})},
\qquad x\in \CC, \quad n\in\Z.
\label{Cgamma}\end{equation}
Formally the asymptotics \eqref{gam2lim2} has been described first in ad hoc manner without derivation in \cite{BMS}.
In the two-dimensional conformal field theory context the variable $b=\sqrt{\omega_1/\omega_2}$ parameterizes
the central charge of the quantum Liouville theory $c_{cft}=1+6(b+b^{-1})^2$. Therefore we deal here with
the limit $c_{cft}\to 1$.

Reflection properties of the ${\bf \Gamma}(x,n)$-function
$$
{\bf \Gamma}(x,-n)=(-1)^n{\bf \Gamma}(x,n), \quad
{\bf \Gamma}(x,n){\bf \Gamma}(-x-2\textup{i},n)=1,
$$
are useful for computations, as well as the difference equations
$$
{\bf \Gamma}(x-\textup{i},n+1)=\frac{n+\textup{i}x}{2} {\bf \Gamma}(x,n),\quad
{\bf \Gamma}(x-\textup{i},n-1)=\frac{n-\textup{i} x}{2} {\bf \Gamma}(x,n).
$$

As seen from \eqref{gam2lim2}, for a special choice of modular parameters
the function $\gamma^{(2)}(u;\omega_1,\omega_2)$
blows up around a special discrete set of points of the argument. As a result, the hyperbolic integrals start to
blow up in this limit as well.
Let us parameterize variables $z$, $\omega_{1,2},$ and $g_j$ in formula \eqref{bi} in the following way
\be\label{zbi1}
z=\textup{i}\sqrt{\omega_1\omega_2}(n+y\delta), \quad
g_j=\textup{i}\sqrt{\omega_1\omega_2}(n_j+s_j\delta),  \quad \sqrt{\omega_1\over \omega_2}=\textup{i}+\delta,
\ee
where  $n, n_j\in\Z+\nu,\; \nu=0, \frac{1}{2},\; y, s_j\in\CC$,
and consider the limit $\delta\to 0^+$ (i.e. $\omega_1/\omega_2\to -1$). First,
the balancing condition \eqref{mu8} reduces to the following constraints for $n_j$ and $s_j$:
\be\label{mu84}
\sum_{j=1}^8 s_j=-4\textup{i}, \qquad \sum_{j=1}^8 n_j=0.
\ee
Second, applying indicated parametrization to the function $I_h$, we find in the limit $\delta\to 0^+$
the diverging asymptotics
\be
I_h(\underline{g};\omega_1,\omega_2)\stackreb{=}{\delta\to 0^+}
\frac{1}{(4\pi\delta)^5} {\bf F}(\underline{s},\underline{n}),
\label{IF}\ee
where ${\bf F}$ is the general complex analogue of the Euler-Gauss $_2F_1$-hypergeometric function
introduced first in the indicated way in \cite{Sarkissian:2020ipg} and in a different way in \cite{DM2019}
\be
{\bf F}(\underline{s},\underline{n})=\frac{1}{8\pi} \sum_{n\in\Z+\nu}\int_{-\infty}^{\infty}(y^2+n^2)
\prod_{k=1}^8{\bf \Gamma}(s_k\pm y,n_k\pm n) dy,
\label{F}\ee
where, as usual, we denoted ${\bf \Gamma}(s\pm y,m\pm n):={\bf \Gamma}(s+y,m+n){\bf \Gamma}(s-y,m-n)$.
For Im$(s_k)<0$ (which corresponds to Re$(g_k)>0$) the contour of integration is $\mathbb{R}$. Otherwise
it is a contour separating arithmetic sequences of poles going to $-\textup{i}\infty$ and $+\textup{i}\infty$
starting from the points $s_k$ and $-s_k$, respectively.

We would like now to derive a difference equation for this function as a limit of the previous
hyperbolic level identity \eqref{br}. Using relations
\begin{eqnarray*} &&
\textup{i}\sqrt{\omega_1\omega_2}(n+y\delta)+\omega_2=\textup{i}\sqrt{\omega_1\omega_2}(n-1+(y-\textup{i})\delta)+O(\delta^2),
\\ &&\qquad  \quad
{\textup{i}\over \omega_1}\sqrt{\omega_1\omega_2}(n+y\delta)=n+\delta(\textup{i}n+y)+O(\delta^2),
\end{eqnarray*}
in the limit $\delta\to 0^+$ we obtain
\be\label{brn1}
{\mathcal B}(\underline{\alpha})({\mathcal F}(s_6-\textup{i},n_6-1,s_7+\textup{i},n_7+1)
-{\mathcal F}(\underline{s},\underline{n}))+(s_6, n_6\leftrightarrow s_7,n_7)+{\mathcal F}(\underline{s},\underline{n})=0,
\ee
where ${\mathcal B}(\underline{\alpha})$ is the potential \eqref{Bgen} with the parametrization
$$
\alpha_k=\frac{1}{2}(\textup{i}s_k-n_k)
$$
and
\be
{\mathcal F}(\underline{s},\underline{n})={{\bf F}(\underline{s},\underline{n})\over {\bf \Gamma}(s_6\pm s_8, n_6\pm n_8){\bf \Gamma} (s_7\pm s_8, n_7\pm n_8)}.
\label{calF}\ee
A striking feature of the derived equation consists in the fact that it cannot be considered as an
analytic difference equation in terms of the complex variables $\alpha_k$, i.e. it cannot be written
in the form \eqref{rateq1}!

Let us take now the second equation for the original hyperbolic $I_h$-function \eqref{br2}.
Applying the same $\delta\to 0^+$ limit we find
\be\label{brn1'}
{\mathcal B}(\underline{\alpha})({\mathcal F}(s_6-\textup{i},n_6+1,s_7+\textup{i},n_7-1)
-{\mathcal F}(\underline{s},\underline{n}))+(s_6, n_6\leftrightarrow s_7,n_7)+{\mathcal F}(\underline{s},\underline{n})=0,
\ee
which is different from the previous equation since ${\mathcal B}(\underline{\alpha})$ is again
the potential  \eqref{Bgen} with a different identification
\begin{equation}
\alpha_k=\frac{1}{2}(\textup{i}s_k+n_k).
\label{iak}\end{equation}
 Jointly the equations \eqref{brn1} and \eqref{brn1'} show that the split
of the variables $\alpha_k$ into two parts --- the continuous and discrete ones is fully justified.
Summarizing, although the original equation \eqref{rateq1} looks as an analytic difference
equation, in our context it should be written in the form \eqref{brn1}, i.e. as a mixture of an
analytic difference equation and a discrete recurrence relation for some supposedly unknown function
${\mathcal F}(\underline{s},\underline{n})$, which is a very unusual picture. The complementary equation,
\eqref{brn1'}, which cannot be written in an ad hoc way, makes such an interpretation inevitable.

Let us remark that special cases of the recurrence-difference equations \eqref{brn1} and \eqref{brn1'}
are related to $6j$-symbols for the Lorentz group (or $SL(2,\CC)$), which have certain applications
in the quantum field theory \cite{DSS}.

\section{Third rational degeneration}

Now we describe a different limit from the hyperbolic level to the rational one.
Let us set
\begin{equation}\label{om1om22}
\sqrt{\omega_1\over \omega_2}=1+\textup{i}\delta, \quad \delta\to 0^+,
\end{equation}
which assumes that $\omega_1+\omega_2=2\sqrt{\omega_1\omega_2}+O(\delta^2)$.
From the quantum Liouville theory point of view this limit corresponds to the central charge value $c_{cft}\to 25$.
 Again, for a special choice of the
argument the $\gamma^{(2)}$-function has a singular behaviour derived in \cite{Sarkissian:2020ipg}
\begin{equation}
\gamma^{(2)}(\sqrt{\omega_1\omega_2}(n+y\delta);\omega_1,\omega_2)\underset{\delta\to 0^+}{=} {\rm e}^{-\frac{\pi \textup{i}}{2}(n-1)^2}
(4\pi\delta)^{n-1}\left(1-\frac{n+\textup{i}y}{2}\right)_{n-1},
\label{limit2}\end{equation}
where $n\in\Z$, $y\in\CC$, $(a)_n$ is the standard Pochhammer symbol: $(a)_0=1$ and
\[
(a)_n=\frac{\Gamma(a+n)}{\Gamma(a)}=
\begin{cases}
 a(a+1)\cdots(a+n-1), & \text{for} \ n>0,\quad \\
\dfrac{1}{(a-1)(a-2)\cdots(a+n)}, &\text{for} \ n<0.
\end{cases}
\]

In order to apply this limit to the hyperbolic integral \eqref{bi}, we parameterize corresponding integration
variable $z$ and the parameters $g_j$ in the following way
$$
z=\sqrt{\omega_1\omega_2}(n+y\delta),  \qquad
g_j=\sqrt{\omega_1\omega_2}(n_j+s_j\delta),
$$
where $y, s_j\in \CC$ and $n, n_j\in\Z+\nu,\; \nu=0, \frac{1}{2}$.
The balancing condition \eqref{mu8}  in the limit $\delta\to 0^+$ yields the constraints
\be\label{mu84'}
\sum_{j=1}^8 s_j=0, \qquad \sum_{j=1}^8 n_j=4.
\ee
Using relation \eqref{limit2}  one finds  \cite{Sarkissian:2020ipg} the asymptotic expression of the integral \eqref{bi}
\be
I_h(\underline{g})\stackreb{=}{\delta\to 0^+} \frac{\textup{i}}{(4\pi\delta)^5} {\bf R}(\underline{s},\underline{n}),
\label{rat2}\ee
where
\be
{\bf R}(\underline{s},\underline{n})=\frac{1}{8\pi}\sum_{n\in \mathbb{Z}+\nu}\int_{-\textup{i}\infty}^{\textup{i}\infty}(y^2+n^2)
\prod_{j=1}^8\left(1-\frac{n_k+\textup{i} s_k\pm (n+\textup{i}y)}{2}\right)_{n_k\pm n-1}\, dy,
\label{rat}\ee
where the contour of integration separates poles appearing from the integrand factors with different choices of the $\pm$ signs. Since the integrands in \eqref{rat} are rational functions, the integrals can be computed
as finite sums of residues yielding rational functions. Moreover, for sufficiently large $|n|$
all poles of the integrands start to lie from one side of the contour of integration, i.e.
the corresponding integrals vanish. Therefore the function ${\bf R}(\underline{s},\underline{n})$
\eqref{rat} is a rational function of its arguments \cite{GSVSfaa}.

Let us consider the limit of interest for the difference equation \eqref{br}. We need the relations
$$
\sqrt{\omega_1\omega_2}(n+y\delta)+\omega_2=\sqrt{\omega_1\omega_2}(n+1+(y-\textup{i})\delta)+ O(\delta^2),
$$
$$
{1\over \omega_1}\sqrt{\omega_1\omega_2}(n+y\delta)=n+\delta(y-\textup{i}n) + O(\delta^2).
$$
With their help we obtain the equation
\be\label{brnrat1}
{\mathcal B}(\underline{ \alpha})(\mathcal{R}(s_6-\textup{i},n_6+1,s_7+\textup{i},n_7-1)-\mathcal{R}(\underline{s},\underline{n}))
+(s_6, n_6\leftrightarrow s_7,n_7)+\mathcal{R}(\underline{s},\underline{n})=0,
\ee
where ${\mathcal B}(\underline{\alpha})$ is the potential from \eqref{brn1'}
with the same identification $\alpha_k=\frac{1}{2}(\textup{i}s_k+n_k)$,
i.e. we obtain the same equation, but for a different function
\be
\mathcal{R}(\underline{s},\underline{n})=\frac{{\bf R}(\underline{s},\underline{n})}
{\prod_{k=6,7}\left(1-\frac{n_k+\textup{i}s_k\pm(n_8+\textup{i} s_8)}{2}\right)_{n_k\pm n_8-1}}.
\label{brn1rat2}\ee
The fact that we have three different solutions for substantially equivalent equations  shows that
the form of solutions depends on the interpretation whether we consider it as a plain analytic
difference equation or a mixed difference-recurrence equation.

Now we apply the limit \eqref{om1om22} to the second hyperbolic equation \eqref{br2}.
Repeating similar steps we find the following new equation
\be\label{brnrat2}
{\mathcal B}(\underline{\alpha})(\mathcal{R}(s_6+\textup{i},n_6+1,s_7-\textup{i},n_7-1)-\mathcal{R}(\underline{s},\underline{n}))
+(s_6, n_6\leftrightarrow s_7,n_7)+\mathcal{R}(\underline{s},\underline{n})=0,
\ee
for the same potential as before ${\mathcal B}(\underline{\alpha})$ with yet another identification
$$
\alpha_k=\frac{1}{2}(-\textup{i}s_k+n_k).
$$
Equation \eqref{brnrat2} differs from \eqref{brn1} by the change of signs in front of the additive 1's
in the potential, by the balancing condition,
as well as the replacement of difference recurrences $s_6-\textup{i},n_6-1$ by $s_6+\textup{i},n_6+1$
(with a similar change for $s_7, n_7$). Thus, we have found three different solutions (given by
${\mathcal I}_{r}$-, ${\mathcal F}$-, and $\mathcal R$-functions) of structurally equivalent
equations  \eqref{rateq1}, \eqref{brn1'}, \eqref{brnrat1}, which are distinguished by the
second relations they satisfy: \eqref{ratid1}, \eqref{brn1}, and \eqref{brnrat2}, respectively.

\section{New rational difference integrable $N$-body problems}

In order to reduce the Ruijsenaars and van Diejen elliptic integrable systems
to the level of complex hypergeometric functions, we describe first such a
degeneration for the symmetric form of elliptic hypergeometric equation \eqref{n1'}.
We use the parametrization
 \begin{equation}
\ve_a=e^{-2\pi v g_a}, \qquad z=e^{-2\pi v y},\qquad p=e^{-2\pi v\omega_1}, \qquad q=e^{-2\pi v\omega_2},
\label{newvepar}\end{equation}
take the limit $v\to 0^+$, and obtain the hyperbolic hypergeometric equation
\begin{eqnarray}\nonumber &&
\frac{\prod_{j=1}^8\sin\frac{\pi}{ \omega_1}(g_j+y)}{\sin{2y\pi\over \omega_1}\sin{(2y+\omega_2)\pi\over \omega_1}} (\chi(y+\omega_2)-\chi(y))
+ \frac{\prod_{j=1}^8\sin\frac{\pi}{ \omega_1}(g_j-y)}{\sin{2y\pi\over \omega_1}\sin{(2y-\omega_2)\pi\over \omega_1}}(\chi(y-\omega_2)-\chi(y))
\\ && \makebox[6em]{}
+\prod_{k=1}^6\sin\frac{\pi}{ \omega_1}\left(g_k +g_7\right)\chi(y)=0,
\label{symmhyper}\end{eqnarray}
with $\sum_{j=1}^8 g_j=2(\omega_1+\omega_2),\, g_8=g_7+\omega_2$.
The limiting form of the function \eqref{f1} solving \eqref{symmhyper} is
\be
\chi(y)={I_h(\omega_2-g_1-C,\ldots,\omega_2-g_5-C,C+y,C-y,C-g_8)\over \gamma^{(2)}(g_8\pm y,  g_6\pm y-4\omega_1; \omega_1,\omega_2)},
\label{chi}\ee
with $C={1\over 2}(g_6+g_8-4\omega_1)$ and the function $I_h$ fixed in  \eqref{bi}.

Let us substitute in the definition \eqref{chi} the parametrization
\be\label{zbi1new}
y=\textup{i}\sqrt{\omega_1\omega_2}(m+u\delta), \quad x=\textup{i}\sqrt{\omega_1\omega_2}(n+v\delta),\quad
g_j=\textup{i}\sqrt{\omega_1\omega_2}(r_j+\gamma_j\delta),
\ee
where $u, v, \gamma_j\in\CC$ and we assume that the discrete variables $m, n, r_j$ can be either integers
or half integers---their admissible values will be determined precisely later on.
Now we take the limit $\sqrt{\frac{\omega_1}{\omega_2}}=\textup{i}+\delta$, $\delta\to 0^+$.
Then we obtain from the relation $g_8=g_7+\omega_2$ the restrictions
\be
r_8=r_7-1, \qquad \gamma_8=\gamma_7-\textup{i},
\ee
and the balancing condition reduces to two constraints
\be
\sum_{j=1}^8 \gamma_j=-4\textup{i}, \qquad \sum_{j=1}^8 r_j=0.
\ee
As to the function $\chi(y)$, we obtain the asymptotic relation
$$
\chi(y)\stackreb{=}{ \delta \to 0^+}
\frac{e^{-\textup{i}\pi(r_8^2+(r_6-4)^2+2m^2)}}{ (4\pi\delta)^{2\textup{i}(\gamma_6+\gamma_8)-7}}
\Psi(\underline{r},\underline{\gamma},m,u),
$$
\begin{equation}
\Psi(\underline{r},\underline{\gamma},m,u)={S(\underline{r},\underline{\gamma},m,u)\over {\bf \Gamma}(\gamma_8\pm u, r_8\pm m){\bf \Gamma} (\gamma_6+4\textup{i}\pm u, r_6-4\pm m)},
\label{calFF}\ee
where $S$-function coincides with the ${\bf F}$-function \eqref{F} under the different
parametrization
\be
S(\underline{r},\underline{\gamma},m,u)=\frac{1}{8\pi} \sum_{n\in\Z+\nu}\int_{-\infty}^{\infty}(n^2+v^2) \prod_{j=1}^8{\bf \Gamma}(s_k\pm v, n_k\pm n)dv,
\ee
with
\begin{eqnarray}\nonumber &&
s_k=-3\textup{i}-\gamma_k-{1\over 2}(\gamma_6+\gamma_8),\quad n_k=1-r_k-{1\over 2}(r_6+r_8),\quad k=1,\ldots, 5,
\\  \nonumber && \makebox[2em]{}
s_{6,7}={1\over 2}(\gamma_6+\gamma_8)+2\textup{i}\pm u,\quad n_{6,7}={1\over 2}(r_6+r_8)-2\pm m,
\\ && \makebox[3em]{}
s_{8}={1\over 2}(\gamma_6-\gamma_8)+2\textup{i},\quad n_{8}={1\over 2}(r_6-r_8)-2.
\label{sr}\end{eqnarray}
One can check that the ${\bf F}$-function balancing condition \eqref{mu84}
is preserved. It is the conditions $n_k\pm n\in\Z$, or $n, n_k\in\Z+\nu,\, \nu=0, 1/2,$ that determine
the quantization rules for the
variables $r_k$ whose values were not completely determined so far. One can see that much depends on the values
of the difference $r_6-r_8$. If this difference is an even integer, then $\nu=0$ and we can set
$r_j, m \in \Z+\mu$ for all $j=1,\ldots, 8$, and some new parameter $\mu=0,1/2$.
If $r_6-r_8$ is an odd integer, then $\nu=1/2$ and, again, we can set all $r_j, m \in \Z+\mu,$ $\mu=0, 1/2$.

Finally, the $S_6$-permutation group symmetric form of the general complex hypergeometric equation is
\begin{eqnarray}\label{n2ratsymm} &&
\frac{\prod_{j=1}^8(\beta_j+z)}{2z(2z+1)}  (\Psi(u-\textup{i},m-1)-\Psi(u,m))
\\ && \makebox[2em]{}
+ \frac{\prod_{j=1}^8(\beta_j-z)}{2z(2z-1)} (\Psi(u+\textup{i},m+1)-\Psi(u,m))
+\prod_{k=1}^6\left(\beta_k+\beta_7\right)\Psi(u,m)=0,
\nonumber\end{eqnarray}
where the variables $z$ and $\beta_k$ are related to $u, m, \gamma_k$ and $r_k$ as follows
$$
\beta_j=\frac{\textup{i}\gamma_j-r_j}{2}, \qquad z=\frac{\textup{i}u-m}{2},
\quad r_j, m \in \Z+\mu, \; \mu=0, \frac{1}{2},
$$
with the additional constraints $r_8=r_7-1, \, \gamma_8=\gamma_7-\textup{i}$.
If we replace $\Psi(u,m)$ by $\Psi(z)$, $\Psi(u-\textup{i},m-1)$ by $\Psi(z+1)$,
and $\Psi(u+\textup{i},m+1)$ by $\Psi(z-1)$ in equation \eqref{n2ratsymm}, then it identically coincides
with the symmetric form of equation \eqref{rateq1}, which emerged from the first rational
degeneration of the hyperbolic integrals, i.e. $\omega_1\to 0$ for fixed $\omega_2$.
If we assume that the latter equation is an analytic difference equation and look for its solutions meromorphic in $z$
(this is a strong restriction on the space of solutions) then one solution is given by the function
\eqref{solution1} after expressing the parameters $\alpha_k$ in terms of $\beta_k$.

However, if we abandon the preassumption that equation \eqref{n2ratsymm} is an analytic difference
equation in $z$-variable, then we get quite different solutions, as we just have described.
Moreover, in such situations two different functions \eqref{calF} and \eqref{brn1rat2} satisfy one
equation \eqref{brnrat1}, but they are distinguished by the second difference-recurrence equations satisfied
by them which are given in \eqref{brn1} and \eqref{brnrat2}, respectively. We do not describe
symmetric forms of these equation for brevity, since all of them have similar forms and differ
by the parametrizations of the variables $\beta_k$ and $z$ only.

Let us turn now to the problem of degenerating the Ruijsenaars and van Diejen models to the level
of complex hypergeometric functions.
First, similar to the elliptic case \eqref{n1}, \eqref{n1'} we interpret the difference-recurrence equation \eqref{n2ratsymm} as a special case of the eigenvalue problem for a rational $N=1$ relativistic Hamiltonian.
From its form we deduce a new relativistic rational $N$-body system appearing from particular
degeneration of the balanced elliptic van Diejen model. This model is described by the Hamiltonian
\be
\mathcal{H}_{rat}=\sum_{j=1}^N \left(B_j(\underline{u},\underline{m})(T_{u_j,m_j}-1)
+B_j(-\underline{u},-\underline{m})(T_{u_j,m_j}^{-1}-1)\right),
\label{newratvD}\ee
where the difference-recurrence operators act as follows
$$
T_{u_j,m_j}^{\pm 1}f(\underline{u},\underline{m})=f(\ldots,u_j\mp \textup{i},\ldots,m_j\mp 1,\ldots)
$$
and the potentials have the form
\be
B_j(\underline{u},\underline{m})=
\prod_{k=1\atop k \ne j}^N\frac{(\beta +z_j+z_k)(\beta+z_j-z_k)}{(z_j+z_k)(z_j-z_k)}
\frac{\prod_{l=1}^8(\beta_l+z_j)}{2z_j(2z_j+1)}.
\label{Bj}\ee
Here we assume that
\be
z_j=\frac{\textup{i}u_j - m_j}{2},
 \qquad \beta=\frac{\textup{i} \gamma-r}{2},  \qquad
 \beta_k=\frac{\textup{i} \gamma_k-r_k}{2},
\label{multivars}\ee
where $u_j, \gamma, \gamma_k\in\CC,$ $ r\in\Z$, $m_j, r_k \in\Z+\mu, \mu =0, \frac{1}{2}$,
with the balancing condition
$$
(2N-2)\gamma+\sum_{k=1}^8\gamma_k=-4\textup{i},\qquad (2N-2)r+\sum_{k=1}^8r_k=0.
$$

The standard rational degeneration of the general van Diejen model was considered in \cite{vD2},
which under the balancing condition yields the Hamiltonian \eqref{newratvD} after the
latter replacement, provided the finite-difference step $2\gamma$ in \cite{vD2}
is normalized to 1 (there is a misprint in formula (3.29) in \cite{vD2} --- in the expression
for $w(z)$ a multiplier 1/4 is missing). Note, however, that the additive part of the potential was conjectured
in \cite{vD2} on the basis of computer calculations for $N<6$ and its factorized form was not derived.

Commuting integrals of motion for the standard rational van Diejen model whose finite-difference
shift step was normalized to 1, which guarantee its integrability,
have the following general operator form \cite{vD,KH}
$$
D_r=\sum_{s=0}^r \sum_{\stackreb{J\in\{1,\ldots, N\},\, |J|=s}{\epsilon_j=\pm1, \, j\in J}}
W_{J,\epsilon_J}(\underline{z}) \prod_{k\in J} T_k^{\epsilon_k},\quad r=1,\ldots, N,
$$
for some coefficients $W_{J,\epsilon_J}(\underline{z})$, where $T_j$ are the ordinary finite-difference operators $T_jf(\underline{z})=f(\ldots,z_j+1,\ldots)$.
Integrability of our Hamiltonian \eqref{newratvD} follows from the fact that the action of
operators $T_{u_j,m_j}$ on the functions of $z_j$  is equivalent to that of $T_j$.
Therefore corresponding integrals of motion are obtained from the operators $D_r$ simply by
using the parametrization \eqref{multivars} and replacement of $T_j$ by $T_{u_j,m_j}$,
and, obviously, they commute with each other.

An inner product for the elliptic van Diejen $N$-body problem has the form \cite{spi:thesis,spi:tmf}
\begin{equation} \label{inner}
\langle \varphi, \psi \rangle_e=
\frac{(p;p)_\infty^N(q;q)_\infty^N}{(4\pi \textup{i})^N N!} \int_{\T^N}\Delta(\underline{x},\underline{t})\;
\varphi(\underline{x}) \psi(\underline{x})\, \frac{dx_1}{x_1}\cdots \frac{dx_N}{x_N},
\ee
where
\be
\Delta(\underline{x},\underline{t})=\prod_{1\le j<k\le N}\!
\frac{\Gamma(tx_j^{\pm1}x_k^{\pm1};p,q)}{\Gamma(x_j^{\pm1}x_k^{\pm1};p,q)}
\ \prod_{j=1}^N \frac{\prod_{k=1}^8\Gamma(t_kx_j^{\pm1};p,q)}{\Gamma(x_j^{\pm 2};p,q)}
\label{Delta}\ee
with the only condition $|t|,\ |t_j|<1$.
The density $\Delta$ is positive-definite for $x_j\in\T$ and real $p,q$ or $p=q^*$.
Additionally one has to impose certain constraint onto the parameters of the model.
Namely, either $t,\ t_j\in\R,\ j=1,\ldots,8,$ or $t\in\R$ and some  $t_j$ are
pairwise complex conjugate to each other $t_j=t^*_k$.

Suppose that the density $\Delta(\underline{x},\underline{t})$ and the functions
$\varphi(\underline{x})$ and $\psi(\underline{x})$ do not contain singularities
lying in the $n$-dimensional annulus $|q|\leq |x_j|\leq |q|^{-1}$, $j=1,\ldots,N$.
This implies additional restrictions $|t|, |t_j|<|q|$.
Under these conditions the general van Diejen Hamiltonian $\mathcal{H}$ is Hermitian with
respect to the taken inner product without any balancing condition on the parameters,
$\langle \varphi, \mathcal{ H}\psi \rangle_e =\langle \mathcal{ H}\varphi, \psi \rangle_e.$
One could take in \eqref{inner} complex conjugated functions $\varphi(\underline{x})^*$
instead of $\varphi(\underline{x})$, but for simplicity we assume that $\varphi(\underline{x})$
are real functions which is natural to expect due to the reflection symmetry $x_j\to x_j^{-1}$.
After imposing the balancing condition $t^{2N-2}\prod_{k=1}^8t_k=p^2q^2$ the Hamiltonian factorizes
\eqref{H} and, evidently, the function $\psi(\underline{x})=1$ becomes its eigenfunction with the zero
eigenvalue. The norm of this eigenfunction is given by the $C_N$-root system extension of the $V$-function,
$\langle 1|1\rangle_e=V(\underline{t};p,q; C_N)$ whose $W(E_7)$-group symmetry was established
by Rains in \cite{rai:limits}.

Now one can consider degeneration of the described scalar product to the hyperbolic
and ``complex'' rational levels. Rigorous degeneration to the hyperbolic level yields
$$
\langle \varphi, \psi \rangle_e \stackreb{\to}{v\to 0^+} e^{\frac{\pi N}{12v}(\frac{5}{\omega_1}
+\frac{5}{\omega_2} +\frac{4g(N-1)}{\omega_1\omega_2})} \langle \varphi, \psi \rangle_h,
$$
where
\begin{eqnarray} \nonumber &&
\langle \varphi, \psi \rangle_h=\frac{1}{N!}
\int_{z_j\in \textup{i}\R}\varphi(\underline{z})\psi(\underline{z})\
\prod_{1\le j<k\le N}\! \frac{\gamma^{(2)}(g \pm z_j\pm z_k;\mathbf{\omega})}
{\gamma^{(2)}(\pm z_j\pm z_k;\mathbf{\omega})}
 \\ && \makebox[4em]{} \times
\ \prod_{j=1}^N  \frac{\prod_{k=1}^8\gamma^{(2)}(g_k\pm z_j;\mathbf{\omega})}
{\gamma^{(2)}(\pm 2z_j;\mathbf{\omega})}\, \frac{dz_j}{2\textup{i}\sqrt{\omega_1\omega_2}}.
\label{measure_hyp} \end{eqnarray}
This is the scalar product valid for general hyperbolic degeneration of the van Diejen model.
If one imposes the balancing constraint $(2N-2)g+\sum_{k=1}^8g_k=2(\omega_1+\omega_2)$ then
corresponding Hamiltonian factorizes and $\langle 1, 1 \rangle_h$ describes a normalization
of the constant eigenfunction of the Hamiltonian with zero eigenvalue.

Further reduction of \eqref{measure_hyp} down to the complex hypergeometric level by taking
$\omega_1/\omega_2\to -1$ \eqref{gam2lim2} yields the degeneration
$$
\langle \varphi, \psi \rangle_h\stackreb{\to}{\delta\to 0^+} \frac{\langle \varphi, \psi \rangle_{rat}}
{(4\pi\delta)^{5N+2\textup{i}\gamma N(N-1)} },
$$
where $\langle \varphi, \psi \rangle_{rat}$ is the scalar product at the level of complex hypergeometric  functions
\begin{eqnarray} \nonumber && \makebox[-2em]{}
\langle \varphi, \psi \rangle_{rat}=
\frac{1}{(8\pi)^N N!} \sum_{m_j\in \Z+\mu}
\int_{u_j\in \R}\varphi(\underline{u},\underline{m})\psi(\underline{u},\underline{m})\
\prod_{1\le j<k\le N}\! \frac{{\bf \Gamma}(\gamma \pm u_j\pm u_k, r\pm m_j\pm m_k)}
{{\bf \Gamma}(\pm u_j\pm u_k, \pm m_j\pm m_k)}
\\ && \makebox[2em]{} \times
\ \prod_{j=1}^N  \left[ \prod_{\ell=1}^8{\bf \Gamma}(\gamma_\ell\pm u_j, r_\ell\pm m_j)\right]
(u_j^2+m_j^2)\, du_j, \quad \textrm{Im}(\gamma), \textrm{Im}(\gamma_k)<0,
\label{complexinner}\end{eqnarray}
where $r_\ell\in \Z+\mu,\, \mu=0, \frac{1}{2}$, $r\in\Z$. Again, this is the general measure when there are no
restrictions on the parameters. Under the following balancing condition
\begin{equation}
(2N-2)\gamma+\sum_{\ell=1}^8\gamma_\ell=-4\textup{i}, \quad (2N-2)r+\sum_{\ell=1}^8r_\ell=0
\label{bal1}\end{equation}
van Diejen Hamiltonian reduces to the one given above \eqref{newratvD}. Then,
for $\varphi(\underline{u},\underline{m})=\psi(\underline{u},\underline{m})=1$ the product
$\langle 1, 1\rangle_{rat}$ is a $C_N$-multivariate complex hypergeometric
function --- an analogue of the $V$-function associated with the Hamiltonian eigenfunction
 $\psi(\underline{u},\underline{m})=1$.  One can check
that under the restrictions on the parameters $\textrm{Im}(\gamma), \textrm{Im}(\gamma_k)<-1$
the Hamiltonian \eqref{newratvD} is hermitian with respect to the scalar product  \eqref{complexinner},
$\langle \varphi,\mathcal{H}_{rat} \psi \rangle_{rat} = \langle \mathcal{H}_{rat}\varphi, \psi \rangle_{rat}$.
The measure of the type indicated in \eqref{complexinner} appeared for the first time in the univariate setting
in an analogue of the Plancherel formula for $3j$-symbols of the group $SL(2,\CC)$ constructed
 by Naimark \cite{Naimark} (see there Theorem 1 and formula (115)).

As follows from the results of \cite{vDS,rai:trans}, for $t_7t_8=pq$ the original elliptic scalar
product $\langle 1, 1\rangle_e$ reduces to the elliptic Selberg integral and can be evaluated explicitly \cite{spi:thesis,spi:tmf}
\begin{eqnarray}\nonumber &&  \makebox[-1em]{}
\langle 1, 1\rangle_e\big|_{t_7t_8=pq}=\frac{(p;p)_\infty^N(q;q)_\infty^N}{(4\pi \textup{i})^N N!} \int_{\T^N} \prod_{1\leq j<k\leq N}
\frac{\Gamma(tx_j^{\pm1} x_k^{\pm1};p,q)}{\Gamma(x_j^{\pm1} x_k^{\pm1};p,q)}
\\  && \makebox[1em]{} \times
\prod_{j=1}^N\frac{\prod_{\ell=1}^6\Gamma(t_\ell x_j^{\pm1};p,q)}{\Gamma(x_j^{\pm2};p,q)}\frac{dx_j}{x_j}= \prod_{j=1}^N\Big(\frac{\Gamma(t^j;p,q)}{\Gamma(t;p,q)}
\prod_{1\leq \ell<s\leq 6}\Gamma(t^{j-1}t_\ell t_s;p,q )\Big),
\label{SintB}\end{eqnarray}
where $t^{2N-2}\prod_{\ell=1}^6t_\ell=pq$.

Simplifying this formula to the hyperbolic
 level one finds \cite{vDS2,rai:limits}
 \begin{eqnarray} \nonumber &&
\langle 1, 1\rangle_h={1\over N!}\int_{z_j\in \textup{i}\R}\prod_{1\leq j < k \leq N}{\gamma^{(2)}(g\pm z_j\pm z_k;\mathbf{\omega})\over \gamma^{(2)}(\pm z_j\pm z_k;\mathbf{\omega})}
\prod_{j=1}^N{\prod_1^6\gamma^{(2)}(g_k\pm z_j;\mathbf{\omega})\over \gamma^{(2)}(\pm 2z_j;\mathbf{\omega})}{dz_j\over 2i\sqrt{\omega_1\omega_2}}
\\ &&  \makebox[2em]{}
= \prod_{j=1}^N\left({\gamma^{(2)}(jg;\mathbf{\omega})\over \gamma^{(2)}(g;\mathbf{\omega})}\prod_{1\leq l < s \leq 6}\gamma^{(2)}((j-1)g+g_l+g_s;\mathbf{\omega})\right)
 \label{bethypm}\end{eqnarray}
with the balancing condition
$$
(2N-2)g+\sum_{k=1}^6g_k=\omega_1+\omega_2.
$$

Finally, reducing this formula to the level of complex hypergeometric
functions we obtain the relation
\begin{eqnarray} \nonumber &&
\langle 1, 1 \rangle_{rat}\big|_{\gamma_7+\gamma_8=-2\textup{i}\atop r_7+r_8=0}
=\frac{(-1)^{2\mu N}}{(8\pi)^N N!} \sum_{m_j\in \Z+\mu}\int_{u_j\in \R}
\prod_{1\le j<k\le N}\! \frac{{\bf \Gamma}(\gamma \pm u_j\pm u_k, r\pm m_j\pm m_k)}
{{\bf \Gamma}(\pm u_j\pm u_k, \pm m_j\pm m_k)}
\\ && \makebox[6em]{} \times
\ \prod_{j=1}^N  \left[ \prod_{\ell=1}^6{\bf \Gamma}(\gamma_\ell\pm u_j, r_\ell\pm m_j)\right]
(u_j^2+m_j^2)\, du_j
\label{complexS} \\ && \makebox[0em]{}
= (-1)^{N (2\mu +r\frac{N-1}{2})} \prod_{j=1}^N\frac{{\bf \Gamma}(j\gamma,jr)}{{\bf \Gamma}(\gamma,r)}
\prod_{1\leq \ell<s\leq 6}{\bf \Gamma}((j-1)\gamma+\gamma_\ell+\gamma_s,(j-1)r+r_\ell+r_s),
\nonumber \end{eqnarray}
where $r\in\Z,\,  m_j, r_\ell\in \Z+\mu,\, \mu=0, \frac{1}{2},$ continuous
variables $\gamma, \gamma_\ell\in\CC$, and the balancing condition has changed to
\begin{equation}
(2N-2)\gamma+\sum_{\ell=1}^6\gamma_\ell=-2\textup{i}, \quad (2N-2)r+\sum_{\ell=1}^6r_\ell=0.
\label{bal2}\end{equation}
Since $r$ is an integer, the root of unity multiplier on the right-hand side of \eqref{complexS} is just a sign factor. Formula \eqref{complexS} describes a new generalization of the famous plain hypergeometric Selberg integral
evaluation formula and it would be interesting to find its reduction to the known three-parameters complex Selberg
integral, see survey \cite{FW} and references therein.

As to the rational degeneration of the elliptic Ruijsenaars model to the level associated with the
complex hypergeometric functions, we suggest the following evident key Hamiltonian, having the
same structure as the operator \eqref{HR},
$$
\mathcal{H}_R=\sum_{j=1}^N\prod_{k=1,\neq j}^N \frac{\gamma+z_j-z_k}{z_j-z_k}\,T_{u_j,m_j},
$$
with the operator $T_{u_j,m_j}$ described before.
Since the action of $T_{u_j,m_j}$ on the variables $z_j$ is the same as of the operator $T_j: z_j\to z_j+1$,
one has the same set of integrals of motions as in the original rational model
with the replacement of $T_j$ by $T_{u_j,m_j}$, i.e. we have a completely integrable model.

\section{Conclusions}

In the present paper we have obtained the following new results. Two new rational degenerations
of the elliptic hypergeometric function and corresponding elliptic hypergeometric equation
were discovered (Sections 3 and 4). These types of degenerations were missed in the
earlier considerations of elliptic hypergeometric functions, see e.g. \cite{BRS,vDS2,rai:limits}.
Taking into account that the elliptic hypergeometric integrals define superconformal indices of four-dimensional
supersymmetric field theories \cite{DO,SV}, such degenerations are associated with the reductions of
the original field theories to lower dimensional spaces.

We have found new rational Calogero-Sutherland type models (Section 5), which are appearing from appropriate
rational degenerations of the elliptic Ruijsenaars and van Diejen models. For the case
related to complex hypergeometric functions we explicitly described the (new) measure
with respect to which the corresponding rational Hamiltonian is Hermitian. This result follows from
the degeneration of the Faddeev modular dilogarithm (the hyperbolic gamma function)
to the complex gamma function rigorously established in \cite{Sarkissian:2020ipg}.

We have found a new degeneration of the elliptic Selberg integral \cite{vDS} given by formula
\eqref{complexS}. It corresponds to the Mellin-Barnes type form of some $N$-dimensional integral
over the complex plane, generalizing the known complex Selberg integrals \cite{FW}, since it contains
more than three free parameters. However, it is not known yet how to perform the transition from
formula \eqref{complexS} to a multiple integral over complex plane. Since the Selberg integral
was useful in several other fields beyond the integrable $N$-body systems \cite{FW}, we expect that
our formula will find similar applications in random matrices, special functions, and so on.
Also we would like to mention a new complicated plain hypergeometric identity given by formula \eqref{ratid1}
which may be useful for applications based on the relations for very-well poised $_9F_8$ hypergeometric
series.

Concerning the physical interpretation of new integrable rational $N$-body models --- they emerge as particular
limiting cases of the standard relativistic quantum mechanical systems, which assumes that
their physical meaning follows from the careful consideration of corresponding degenerations.
In particular, one can say that they describe relativistic particles with additional discrete
degrees of freedom related to the spins of two-dimentional conformal field theory or fluxes
of the two-dimensional vortices. For example, similar bilateral infinite sums over the discrete
magnetic charges (fluxes) emerge in the computations of three-dimensional superconformal indices \cite{KSV}.

It would be also interesting to consider the limit from our recurrence-difference relations to the
recurrence-differential equations for the described models. However, such a continuous
limit was not considered even for the standard rational degeneration \cite{vD}, currently only
the reduction to elliptic Inozemtsev model was described with known further degeneration down
to the Calogero model \cite{vD2,R3}. Therefore it is necessary first to investigate the continuous limit of
the standard rational degeneration, which we plan to do in the future work together with
similar analysis for our situation. Moreover, it is necessary to find relations to the complex
differential equations for hypergeometric functions considered in \cite{MN}.
Another point of future interest --- it is necessary to describe degeneration of the
elliptic  Ruijsenaars and van Diejen models associated with the third rational limit of the elliptic
hypergeometric equation \eqref{brnrat1} or \eqref{brnrat2}.

Moreover, there is further extension of the constructions described in this paper. Namely, the
standard elliptic hypergeometric functions can be extended to the ``rarefied'' versions by
adding some discrete parameters associated with the lens space. Respectively, there exist the
corresponding rarefied analogues of the elliptic hypergeometric equation, its solutions
described by the $V$-function and the corresponding solutions of the Yang-Baxter equation
\cite{spi:jphysa}. It should be possible to adapt the limiting procedure described in the present
work to these more general constructions and build, expectedly, some further deformations of the
rational integrable models described above. We hope to consider all the mentioned complementary
problems in future separate works.

\medskip
{\bf Acknowledgments.} This study has been partially funded within the framework
of the HSE University Basic Research Program. The authors are indebted to J. F. van Diejen for
valuable comments and inspiring discussions.

\end{document}